\documentclass[aps,pra,twocolumn,superscriptaddress,longbibliography]{revtex4-2}
\usepackage{amsfonts}
\usepackage{bbding}
\usepackage{amssymb}
\usepackage[title]{appendix}
\usepackage{color}
\usepackage{amsmath}
\usepackage{graphicx}
\usepackage{txfonts}
\usepackage{bm}
\usepackage{natbib}
\usepackage{graphicx}
\usepackage[percent]{overpic}
\usepackage[colorlinks,linkcolor=blue,citecolor=blue,urlcolor=blue]{hyperref}
\usepackage{orcidlink}
\preprint{APS/123-QED}

\DeclareMathAlphabet{\pazocal}{OMS}{zplm}{m}{n}

\begin{document}

\title{Steady-state phase transition in one-dimensional hybrid contact process}

\author{Lin Shang\orcidlink{0009-0007-5840-2351}}
\affiliation{School of Physics, Dalian University of Technology, Dalian 116024, China}

\author{Shuai Geng\orcidlink{0009-0000-8687-0493}}
\affiliation{School of Physics, Dalian University of Technology, Dalian 116024, China}

\author{Xingli Li\orcidlink{0000-0003-2339-6557}}
\affiliation{Department of Physics, The Chinese University of Hong Kong, Shatin, New Territories, Hong Kong, China}
\affiliation{The Chinese University of Hong Kong Shenzhen Research Institute, Shenzhen 518057, China
}
\affiliation{Lanzhou Center for Theoretical Physics, Key Laboratory of Theoretical Physics of Gansu Province, Key Laboratory of Quantum Theory and Applications of MoE, Gansu Provincial Research Center for Basic Disciplines of Quantum Physics, Lanzhou University, Lanzhou 730000, China}

\author{Jiasen Jin\orcidlink{0000-0002-7118-7488}}
\email{jsjin@dlut.edu.cn}
\affiliation{School of Physics, Dalian University of Technology, Dalian 116024, China}

\begin{abstract}
We investigate the steady-state phase transition in a one-dimensional hybrid contact process. We implement the single-site and cluster mean-field approximations based on the effective fields and present all the possible steady states of the system. We show the existence of the stable absorbing and active phases, and the bistable region in the long-time limit. The saddle-node bifurcation is observed at the boundary between the absorbing phase and the bistable region, suggesting a discontinuous phase transition. While the absorbing to active phase transition is continuous. To characterize the nonclassical scaling behavior of the continuous phase transition, we extract the true critical points and exponents by means of the coherent anomaly method.
\end{abstract}

\date{\today}

\maketitle 

\section{Introduction}

In recent years, the nonequilibrium dynamics and phase transitions in open quantum many-body systems have attracted extensive attention from both experimental and theoretical perspectives~\cite{Carusotto2013,Noh2017,Daley2014,breuerTheoryOpenQuantum2007,Rivas2020,Alexandre2013,Lee2011,Fitzpatrick2017,Collodo2019}. 
Compared with classical stochastic systems, open quantum systems exhibit additional features such as the quantum coherence, quantum correlations, entanglement, and quantum fluctuations. These effects can strongly modify the dynamical relaxation, steady-state properties, and critical behavior of the system ~\cite{Carusotto2013,Daley2014,breuerTheoryOpenQuantum2007,Sieberer2025}. 
Recent advances in controllable quantum platforms, including Rydberg atoms, optical lattices, trapped ions, photonic systems, and superconducting circuits, have also provided promising settings for probing nonequilibrium many-body dynamics and dissipative phase transitions~\cite{Baumann2011,Baumann2010,Bloch2012,Gross2017,Browaeys2020,Bernien2017,Zhang2017,Fitzpatrick2017,Collodo2019}.

A paradigmatic example of nonequilibrium phase transitions is the absorbing-state phase transition. In this context, the classical contact process is one of the simplest and most important models. It describes the time evolution of the excitations under competition between the local decay and the branching (coagulation) in a one-dimensional system with nearest-neighbor interactions. The local decay tends to destroy the excitation of each site, while the branching (coagulation) exhibits a correlated activation (deactivation) conditioned on its neighbor being excited \cite{Harris1974,Liggett1985,Mollison2018,Grassberger1983,MarroDickman1999}.
If the local decay dominates, the system will eventually evolve to the so-called absorbing state with vanishing excitations; otherwise, the system will evolve to the active state with non-zero excitations among the sites. The steady state, that is the asymptotic state in the long-time limit, of the system undergoes a continuous phase transition when the controlled parameter crosses the critical point. Moreover, the continuous phase transition from the absorbing to the active state in the classical contact process belongs to the directed percolation universality class \cite{Grassberger1978,Janssen1981,grassbergerPhaseTransitionsSchlogls1982,Hinrichsen2000,Odor2004,Henkel2008}.
Since its introduction, the contact process and its variants have been widely used to simulate the spreading phenomena in epidemic dynamics, population dynamics, social dynamics, and information propagation~\cite{Mollison2018,Kuhr2011,Durrett1994,Claudio2009}. 

The quantum contact process (QCP) is a coherent extension of the classical contact process. The key feature of QCP is that the classical stochastic correlated activation and deactivation are replaced by the coherent operations, in which the coherence may arise in the system
~\cite{Matteo2016,Michael2017,Dietrich2018,Federico2019,Edward2019,Edward2020,Jo2021}. 
It has been shown that the critical properties of the absorbing phase transition in the QCP are modified by the quantum fluctuations ~\cite{Federico2019,Edward2019,Jo2021},
and the strong coherent dynamics may drive the system toward discontinuous behavior or bistability~\cite{Matteo2016,Michael2017,Dietrich2018}. The evidence of the discontinuous phase transition of QCP is shown by a self-consistent effective field method and the metastability of the system is uncovered through the Liouvillian spectrum~\cite{Shang2026}. In addition, the coherent processes can also substantially alter the spreading dynamics of the quantum epidemic model in the open quantum spin system~\cite{PhysRevLett.119.140401}. 

These results show that the role of quantum coherence in absorbing-state phase transition remains subtle and important. It would be interesting to investigate how the continuous nature of phase transition is affected by quantum coherence. Motivated by this issue,  we extend the study in Ref. \cite{Shang2026} to the hybrid contact process (HCP) in one dimensional, in which the coagulation and branching are implemented both quantum and classically. This model interpolates between the QCP and the purely classical contact process. In particular, we ask whether introducing the quantum process changes the critical exponent of steady-state phase transition observed in the classical contact process. By combining the cluster mean-field approximation (CMF), Liouvillian spectral analysis, and the coherent anomaly method (CAM), we characterize the steady-state phase transition and extract the corresponding critical behavior~\cite{Jin2016,Minganti2018,Mori2020,Shirai2023,Suzuki1986,Suzuki1987,Park2005,Jin2021}. 

The paper is organized as follows. In Sec.~\ref{model}, we introduce the setup of the HCP in a one-dimensional spin-1/2 system and the quantum master equation that describing its dynamical evolution. In Sec.~\ref{MF approximation}, we investigate the steady-state behavior of the system under the single-site mean-field approximation by solving the  Bloch equation, and compare the results with those of the QCP. In Sec.~\ref{Liouvillian spectrum}, we discuss the Liouvillian spectrum of the HCP. In Sec.~\ref{Cluster Mean-field approximation}, we incorporate the short-range correlations by means of the CMF
approximation to analyze the steady-state properties, calculate the critical points of the HCP. In Sec.~\ref{Coherent anomaly method}, we extract the critical points and exponents of the continuous phase transition in the HCP with the coherent anomaly method. We summarize in Sec.~\ref{summary}.

\section{The model}
\label{model}

The one-dimensional HCP considered in this paper is defined as follows. The total system is composed of a chain of qubits. Namely, there are two orthogonal states on each site, being either occupied (with an excitation) or empty (without excitation). An occupied site may activate its neighboring empty site or deactivate its neighboring occupied sites. The former is the so-called branching and the latter is the coagulation. Additionally, an occupied site has the possibility to become empty spontaneously, the so-called self-desctruction. In the HCP, the self-destruction of the excitation on each site works incoherently via the local decay at a rate $\Gamma$. The coagulation and branching may take place through both the coherent and incoherent means. The coherent process is implemented via a unitary operation, i.e. a tensor product of the projector and a NOT-gate operator, acting on a pair of nearest neighboring sites. The incoherent process is implemented via the correlated dissipator, in which the action of an activation or deactivation on a site is conditioned on its nearest neighbor site being occupied.

\begin{figure}[htbp]
\centering
\includegraphics[width=1\linewidth]{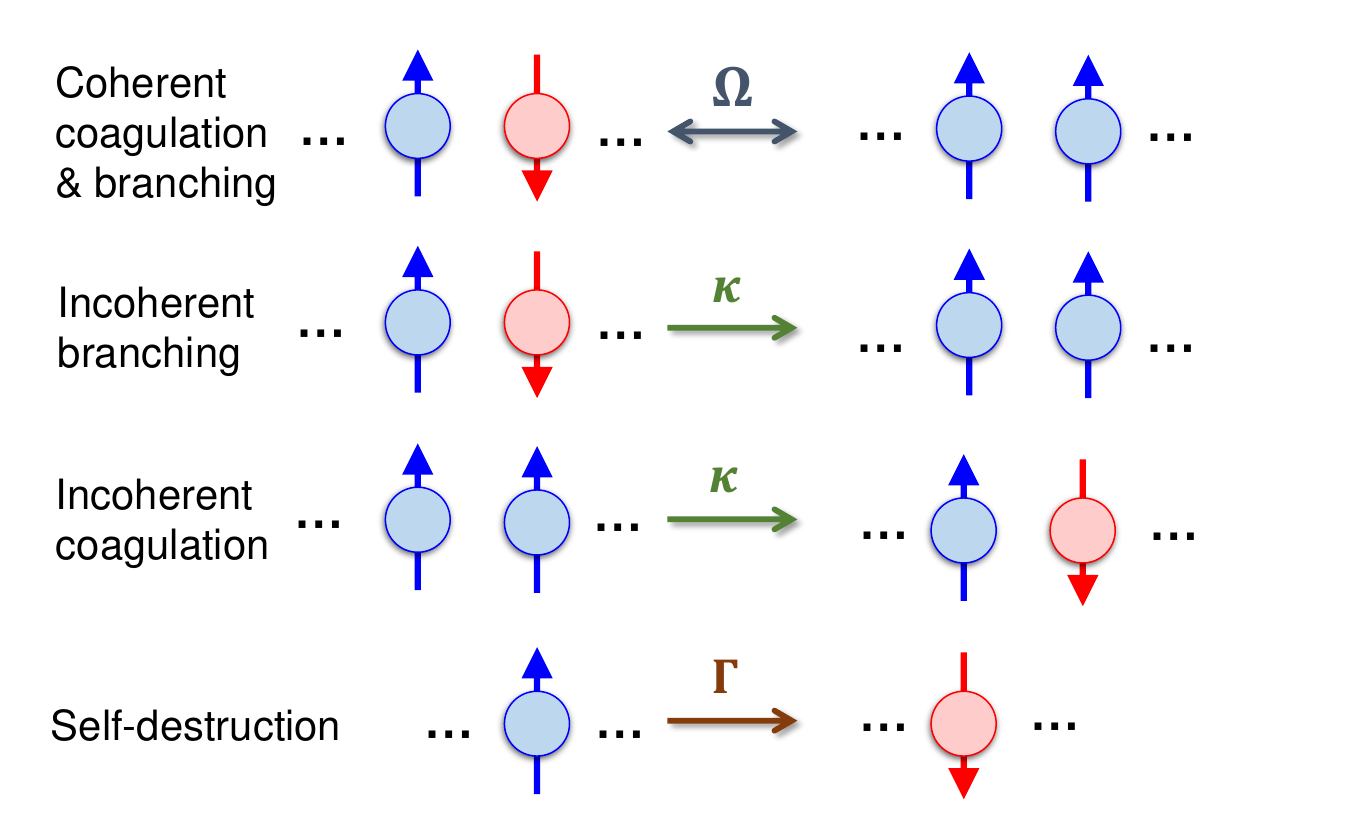}
\caption{Schematic illustration of the coagulation, branching, and self-destruction in one-dimensional hybrid contact process. In the quantum process, when a neighboring sites is occupied, the spin at each site undergoes coherent flipping between the occupied state (spin-up) and the empty state (spin-down) at a rate $\Omega$. Correspondingly, in the classical process, the incoherent activation and deactivation occur at a rate $\kappa$. In addition, each spin undergoes incoherent decay to the empty state at a rate $\Gamma$.}
\label{fig:1}
\end{figure}

To simulate the one-dimensional HCP, we consider a quantum spin-$1/2$ chain with local dissipation acting on each site. The setup is illustrated in Fig.~\ref{fig:1}. For each spin, the occupied state is denoted by $|\uparrow\rangle$, while the empty state is denoted by $|\downarrow\rangle$, where $|\uparrow\rangle$ and  $|\downarrow\rangle$ are the eigenstates of the spin $z$-component.
Under the Markovian approximation, the time evolution of the density matrix is governed by the Lindblad master equation~\cite{breuerTheoryOpenQuantum2007},
\begin{equation}
\frac{\text{d}}{\text{d}t}\rho(t)={\cal L}[\rho]:=-i[\hat{H},\rho]+{\cal D}_{\text{corr}}[\rho]+{\cal D}_{\text{loc}}[\rho],
\label{eq:(2)}
\end{equation}
where the notation $[\cdot,\cdot]$ stands for the commutator and ${\cal L}$ is the Liouvillian superoperator. We set $\hbar=1$ hereinafter. 

The first term on the right-hand side of Eq. (\ref{eq:(2)}) describes the coherent coagulation and branching governed by the Hamiltonian $\hat{H}$,
\begin{equation}
 \hat{H}=\Omega\sum_{j=1}^{L-1}(\hat{\sigma}_{j}^{x}\hat{n}_{j+1}+\hat{n}_j\hat{\sigma}_{j+1}^{x}),
\label{eq:(1)}
\end{equation}
where $\Omega$ denotes the coherent interaction strength and $L$ is the length of the spin chain. Here, $\hat{\sigma}_j^{\alpha}$ $(\alpha=x,y,z)$ are the Pauli operators acting on the $j$-th site, 
$\hat{n}_j=\hat{\sigma}_j^+\hat{\sigma}_j^-$ is the particle number operator, which also serves as the projector onto the occupied state. The raising and lowering operators on the $j$-th site are defined as 
$\hat{\sigma}_j^{\pm}=(\hat{\sigma}_j^x\pm i\hat{\sigma}_j^y)/2$.

The second term on the right-hand side of Eq. (\ref{eq:(2)}) describes the incoherent coagulation and branching, which are implemented through correlated dissipations at a rate $\kappa$. The corresponding correlated Lindbladian superoperators are given as follows,
\begin{equation}
{\cal D}_{\text{corr}}[\rho]=\sum_{\alpha=\pm}\sum_{j=1}^{L}{\left[\hat{L}_j^{\alpha}\rho (\hat{L}_j^{\alpha})^\dagger-\frac{1}{2}\{(\hat{L}_j^{\alpha})^\dagger\hat{L}_j^{\alpha},\rho\}\right]},
\label{eq:(2corr)}
\end{equation}
where the $\{\cdot,\cdot\}$ stands for the anticommutator, and the correlated jump operators are given by $\hat{L}_j^{\pm} = \sqrt{\kappa}\hat{n}_{j'}\hat{\sigma}_j^{\pm}$ with $j'$ being the nearest neighbors of $j$. 

The third term on the right-hand side of Eq. (\ref{eq:(2)}) describes the self-destruction on each site at a rate $\Gamma$, the local Lindbladian is given by
\begin{equation}
{\cal D}_{\text{loc}}[\rho]=\sum_{j=1}^{L}{\left(\hat{L}_j\rho\hat{L}_j^\dagger-\frac{1}{2}\{\hat{L}_j^\dagger\hat{L}_j,\rho\}\right)},
\label{eq:(2loc)}
\end{equation}
where the local jump operator is $\hat{L}_j=\sqrt{\Gamma}\hat{\sigma}_j^-$.

We are interested in the steady state of Eq. (\ref{eq:(2)}) in the thermodynamic limit, i.e. $\rho_{\text{ss}}=\lim_{t\rightarrow\infty}\lim_{L\rightarrow\infty}{\rho(t,L)}$, the subscript `ss' denotes the steady state. It should be noted that the two limits in defining $\rho_{\text{ss}}$ do not commute. We adopt the averaged particle density as the order parameter $\langle \hat{n}\rangle_{\text{ss}}=\sum_{j=1}^L{\langle\hat{n}_j\rangle_{\text{ss}}}/L$. The zero $\langle \hat{n}\rangle_{\text{ss}}$ means that all the spins are pointing down to the $z$-direction and the system is in the so-called \textbf{absorbing state}. The nonzero $\langle \hat{n}\rangle_{\text{ss}}$ means that initial excitations survive after sufficient long time and the system is thus in the so-called \textbf{active state}.

One can see that when the quantum interaction is set $\Omega=0$, the system reduces to the classical contact process. In contrast, when $\kappa=0$, the model reduces to the QCP. In this work, we focus on the steady-state phase diagram and critical behavior of the system when both the classical and quantum interactions are nonzero. In particular, we find that, for small $\Omega$, the system undergoes a continuous phase transition from the absorbing state to the active state as the classical interaction strength $\kappa$ is varied.

\section{Mean-field approximation}
\label{MF approximation}
\subsection{The mean-field master equation and Bloch equations}
In order to get a preliminary impression on the steady states of the HCP, we adopt the single-site mean-field (MF) approximation, which is one of the commonly used methods for solving the steady-state solutions of the open quantum many-body systems~\cite{Lee2011,Jin2013} . The MF approximation is based on the Gutzwiller factorization ansatz, in which the total density matrix of the system $\rho$ is factorized into the following product state,
\begin{equation}
\rho=\bigotimes_{j=1}^L\rho_j, 
\label{eq:GutzwillerFactoriztion}
\end{equation}
where $\rho_j$ is the density matrix of each site and are assumed to be identical, i.e.  $\rho_j = \rho^{\text{mf}}, \forall j$. 

The time evolution of $\rho_j$ can be obtained by substituting Eq. (\ref{eq:GutzwillerFactoriztion}) into Eq. (\ref{eq:(2)}) and take the partial trace over the sites other than $j$, i.e. $\frac{\text{d}}{\text{d}t}\rho_j=\text{tr}_{\ne{j}}\frac{\text{d}}{\text{d}t}\rho$. In this manner, the many-body master equation is reduced to a single-site master equation involving the density matrix of a single site. As a consequence, the interactions between the given site and its nearest neighbors are replaced by the expectation values of the corresponding observables, or say the mean fields. Since the density matrix of each site is identical, the single-site MF master equation yields,
\begin{equation}
\frac{\text{d}}{\text{d}t}\rho^{\text{mf}}=-i[\hat{H}^{\text{mf}},\rho^{\text{mf}}]+{\cal D}_{\text{corr}}^{\text{mf}}[\rho^{\text{mf}}]+{\cal D}_{\text{loc}}^{\text{mf}}[\rho^{\text{mf}}],
\label{mf_mastereqution}
\end{equation}
where $\hat{H}^{\text{mf}}=2\Omega(\langle\hat{\sigma}^x\rangle\hat{n} + \langle\hat{n}\rangle\hat{\sigma}^x)$ represents the MF Hamiltonian. The MF correlated dissipator 
reads ${\cal D}_{\text{corr}}^{\text{mf}}[\rho^{\text{mf}}]=\kappa\left[2\hat{n}\rho^{\text{mf}}\hat{n}+2\langle \hat{n}\rangle(\hat{\sigma}^+\rho^{\text{mf}}\hat{\sigma}^- +\hat{\sigma}^-\rho^{\text{mf}}\hat{\sigma}^+-\rho^{\text{mf}})-\{\hat{n},\rho^{\text{mf}}\}\right]$, and the local dissipator reads
${\cal D}_{\text{loc}}^{\text{mf}}[\rho^{\text{mf}}] = \frac{\Gamma}{2}\left[2\hat{\sigma}^-\rho^{\text{mf}}\hat{\sigma}^+-\{\hat{n},\rho^{\text{mf}}\}\right]$ is the dissipation term. In deriving the master equation (\ref{mf_mastereqution}), we have omitted the subscript $j$ of the density matrix and the observables, because the local observables at each site are identical, i.e., $\langle\hat{O}_j\rangle=\text{tr}(\rho_j\hat{O}_j) = \langle\hat{O}\rangle,\forall j$. 

By virtue of Eq. (\ref{mf_mastereqution}), the system of the single-site MF Bloch equations yields as follows,
\begin{equation}\label{eq:(5)}
\begin{aligned}
\frac{\text{d}\langle\hat{\sigma}^x\rangle}{\text{d}t}
&= -2\Omega\langle\hat{\sigma}^x\rangle\langle\hat{\sigma}^y\rangle
   - \frac{\Gamma}{2}\langle\hat{\sigma}^x\rangle
   - \kappa\langle\hat{\sigma}^x\rangle(2+\langle\hat{\sigma}^z\rangle), \\[4pt]
\frac{\text{d}\langle\hat{\sigma}^y\rangle}{\text{d}t}
&= 2\Omega\langle\hat{\sigma}^x\rangle\langle\hat{\sigma}^x\rangle
   - 2\Omega(\langle\hat{\sigma}^z\rangle+1)\langle\hat{\sigma}^z\rangle
   - \frac{\Gamma}{2}\langle\hat{\sigma}^y\rangle \\
&\quad
   - \kappa\langle\hat{\sigma}^y\rangle(2+\langle\hat{\sigma}^z\rangle), \\[4pt]
\frac{\text{d}\langle\hat{\sigma}^z\rangle}{\text{d}t}
&= 2\Omega(\langle\hat{\sigma}^z\rangle+1)\langle\hat{\sigma}^y\rangle
   - \Gamma(\langle\hat{\sigma}^z\rangle + 1)
   - 2\kappa\langle\hat{\sigma}^z\rangle(1+\langle\hat{\sigma}^z\rangle).
\end{aligned}
\end{equation}

\subsection{The steady-state solutions and stability}
The steady-state solution to Eqs. (\ref{eq:(5)}) can be obtained by setting $\text{d}\langle\bm{\hat{\sigma}}\rangle/\text{d}t=0$. Some of the steady-state solutions are nonphysical because the norm of the corresponding Bloch vector is larger than 1, in other words, the particle number is negative. Discarding the nonphysical solutions, one can obtain three sets of real steady-state solutions $\langle\bm{\hat{\sigma}}\rangle_{\text{ss}}=(\langle\hat{\sigma}^x\rangle_{\text{ss}},\langle\hat{\sigma}^y\rangle_{\text{ss}},\langle\hat{\sigma}^z\rangle_{\text{ss}})$ as follows,
\begin{equation}
\langle\hat{\bm{\sigma}}\rangle_{\text{ss},0}=(0,0,-1),
\label{eq:mf_sol_0}
\end{equation}
and 
\begin{equation}
\langle\hat{\bm{\sigma}}\rangle_{\text{ss},\pm}=\left(0,\frac{\Gamma+2\kappa\langle\hat{\sigma}^z\rangle_{\text{ss},\pm}}{2\Omega},\langle\hat{\sigma}^z\rangle_{\text{ss},\pm}\right),
\label{eq:mf_sol_pm}
\end{equation}
where
\begin{equation}
\langle\hat{\sigma}^z\rangle_{\text{ss},\pm} = \frac{-\omega^2-\kappa(\Gamma+\kappa)\pm\sqrt{2\omega^2(2\kappa^2+\Omega^2)-2\Omega^2(\Gamma+\kappa)^2}}{2\omega^2},
\end{equation}
with $\omega^2 = 2\Omega^2+\kappa^2$.

Moreover, the stabilities of the physical acceptable solutions (\ref{eq:mf_sol_0}) and (\ref{eq:mf_sol_pm}) are determined by the eigenvalues of the Jacobian matrix $\bm{M}$. The entry of the Jacobian matrix is  $\bm{M}_{\alpha\beta}:=\partial{f_\alpha}/\partial\langle\hat{\sigma}^\beta\rangle$, where $f_\alpha=\text{d}\langle\hat{\sigma}^\alpha\rangle/\text{d}t$ with $\alpha,\beta=x,y,z$. The Jacobian matrix $\bm{M}$ is thus given as follows,

\begin{widetext}
\begin{equation}\label{eq:(6)}
    \bm{M} =
    \begin{pmatrix}
      -2\Omega\langle\hat{\sigma}^y\rangle - \frac{\Gamma}{2} -\kappa(2+\langle\hat{\sigma}^z\rangle) &
      -2\Omega\langle\hat{\sigma}^x\rangle &
      -\kappa\langle\hat{\sigma}^x\rangle \\
      4\Omega\langle\hat{\sigma}^x\rangle &
      -\frac{\Gamma}{2}-\kappa(2+\langle\hat{\sigma}^z\rangle) &
      -2\Omega(1+2\langle\hat{\sigma}^z\rangle) - \kappa\langle\hat{\sigma}^y\rangle \\
      0 &
      2\Omega(\langle\hat{\sigma}^z\rangle + 1) &
      2\Omega\langle\hat{\sigma}^y\rangle - \Gamma-2\kappa(1+2\langle\hat{\sigma}^z\rangle)
    \end{pmatrix}.
\end{equation}
\end{widetext}

Substituting Eq. (\ref{eq:mf_sol_0}) or (\ref{eq:mf_sol_pm}) into Eq. (\ref{eq:(6)}), one obtains the Jacobian matrix in steady state. The presence of an eigenvalue with a positive real part indicates that the solution is unstable. Conversely, a steady-state solution is stable if all its eigenvalues possess negative real parts. This enables us to find out the all the stable steady states and, consequently, to identify the steady-state phases. Actually, it turns out that the physical acceptable solution $\langle\hat{\bm{\sigma}}\rangle_{\text{ss},-}$ is always unstable. 

\subsection{The MF phase diagram}
\begin{figure}[t!]
\centering
\includegraphics[width=0.9\linewidth]{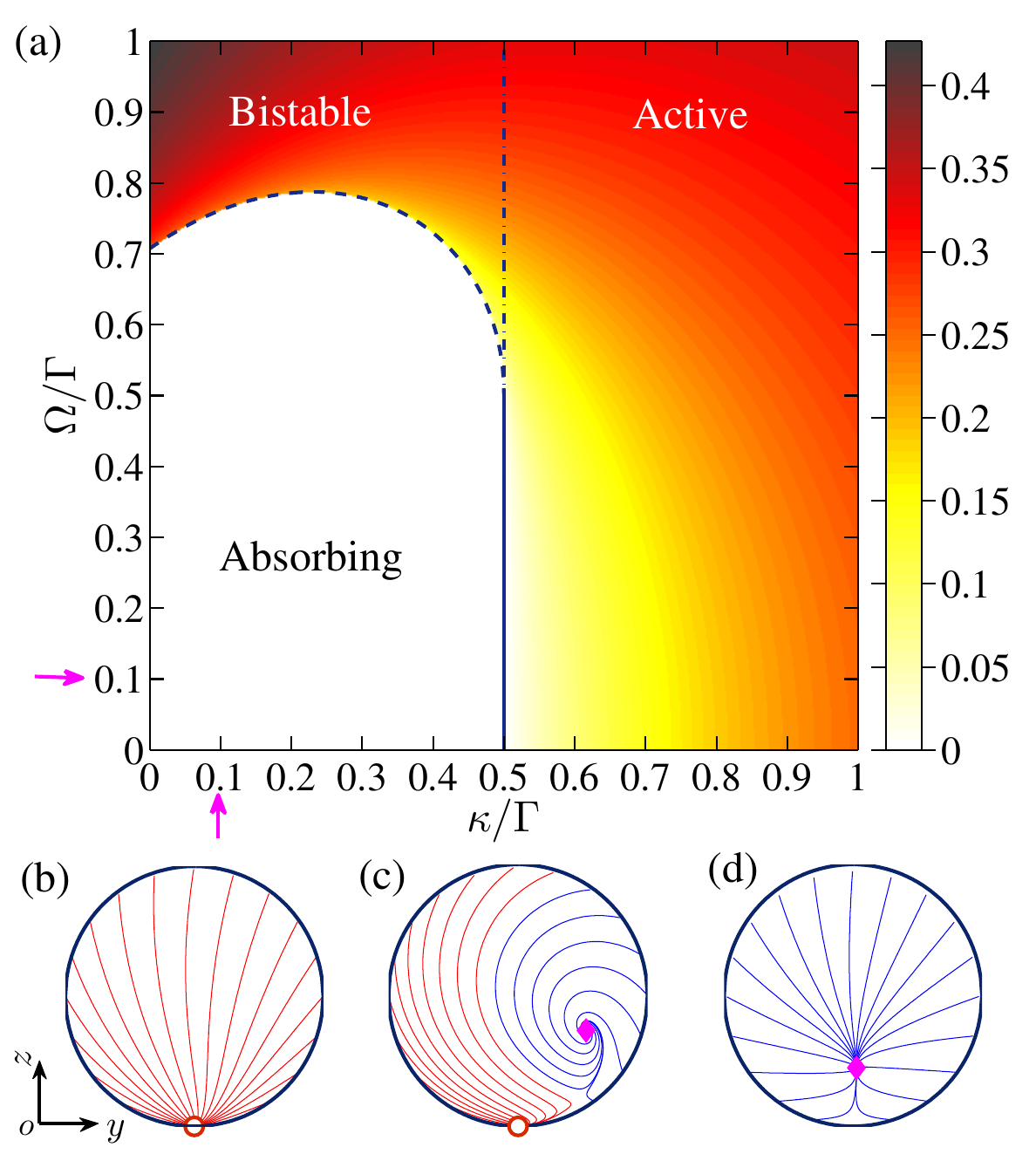}
\caption{(a) The mean-field steady-state phase diagram of HCP in the $\Omega-\kappa$ plane. The white region enclosed by the solid and dashed lines represents the absorbing phase with zero particle density, associated with the steady-state solution $\langle\hat{\bm{\sigma}}\rangle_{\text{ss},0}$. The colored region to the right of the solid and dashed-dotted lines, characterized by $\kappa/\Gamma=1/2$, represents the active phase with nonzero particle density and it is associated to the steady-state solution $\langle\hat{\bm{\sigma}}\rangle_{\text{ss},+}$. The region enclosed by the dashed and dashed-dotted lines represents the bistable region, in which the system could evolve to either the absorbing or the active state in the long-time limit depending on the initial state. The color bar denotes the steady-state particle density $\langle\hat{n}\rangle_{\text{ss}}$. The trajectories of the time evolution with various initial state on the $y-z$ plane of the Bloch sphere in the absorbing phase (b), bistable region (c) and active phase (d). The initial state is characterized by $|\psi\rangle=\cos{(\theta/2)}|\uparrow\rangle\pm i\sin{(\theta/2)}|\downarrow\rangle$ where $\theta$ is the polar angle. The red lines represent the trajectories attracted by the the absorbing state $\langle\bm{\hat{\sigma}}\rangle_{\text{ss},0}$ (the red empty circle) and  the blue lines represent the trajectories attracted by the active state $\langle\bm{\hat{\sigma}}\rangle_{\text{ss},+}$ (filled magenta diamond). The parameters are chosen as $(\Omega,\kappa)/\Gamma=(0.1,0.1)$, $(0.9,0.1)$, and $(0.1,0.9)$ for (b), (c) and (d), respectively.
} 
\label{fig:0}
\end{figure}

In Fig. \ref{fig:0}, we show the MF phase diagram of the HCP in the $\Omega-\kappa$ plane. Recall that the order parameter is chosen as the steady-state particle density which can be computed as $\langle\hat{n}\rangle_{\text{ss}}=(1+\langle\hat{\sigma}^z\rangle_{\text{ss}})/2$. It can be seen that the phase diagram consists of three regions: the absorbing phase, the active phase and the bistable region in which the long-time limit particle density depends on the initial state.

The absorbing phase, enclosed by the solid and dashed lines in Fig. \ref{fig:0}(a), is indicated by $\langle\hat{n}\rangle_{\text{ss}}=0$. It is associated with the solution $\langle\bm{\hat{\sigma}}\rangle_{\text{ss},0}=(0,0,-1)$. The maximal real part of the eigenvalues of the corresponding Jacobian matrix is given by $\text{Re}(\lambda_0)/\Gamma=\max{\{2\kappa/\Gamma-1,-\kappa/\Gamma-1/2\}}$. One can find that $\langle\bm{\hat{\sigma}}\rangle_{\text{ss},0} $
is stable when $\kappa/\Gamma < 1/2$ and is independent of $\Omega$.
As $\kappa/\Gamma$ exceeds $1/2$, the absorbing state becomes unstable and the active steady-state phase with nonzero particle density emerges, lying to the right of the solid and dash-dotted lines in Fig. \ref{fig:0}(a). It is associated with the steady-state solution $\langle\bm{\hat{\sigma}}\rangle_{\text{ss},+}$. In addition, the bistable region is enclosed by the dashed and the dash-dotted line in Fig. \ref{fig:0}(a). 

In Figs. \ref{fig:0}(b)-(d), the time-evolution trajectories of the Bloch vector on the $y-z$ plane are shown. Starting from various initial states $|\psi\rangle=\cos{(\theta/2)}|\uparrow\rangle+e^{i\phi}\sin{(\theta/2)}|\downarrow\rangle$, with the azimuthal angle $\phi=\pm\pi/2$, the system is always attracted by a unique steady state in the absorbing and active phases. In the bistable region, however, there are two attractors in the $y-z$ plane characterized by $\langle\bm{\hat{\sigma}}\rangle_{\text{ss},0}$ and $\langle\bm{\hat{\sigma}}\rangle_{\text{ss},+}$. Therefore the final state in the long-time limit depends on the initial state.

\subsection{The discontinuous and continuous phase transitions}

\begin{figure}[t!]
\centering
\includegraphics[width=0.9\linewidth]{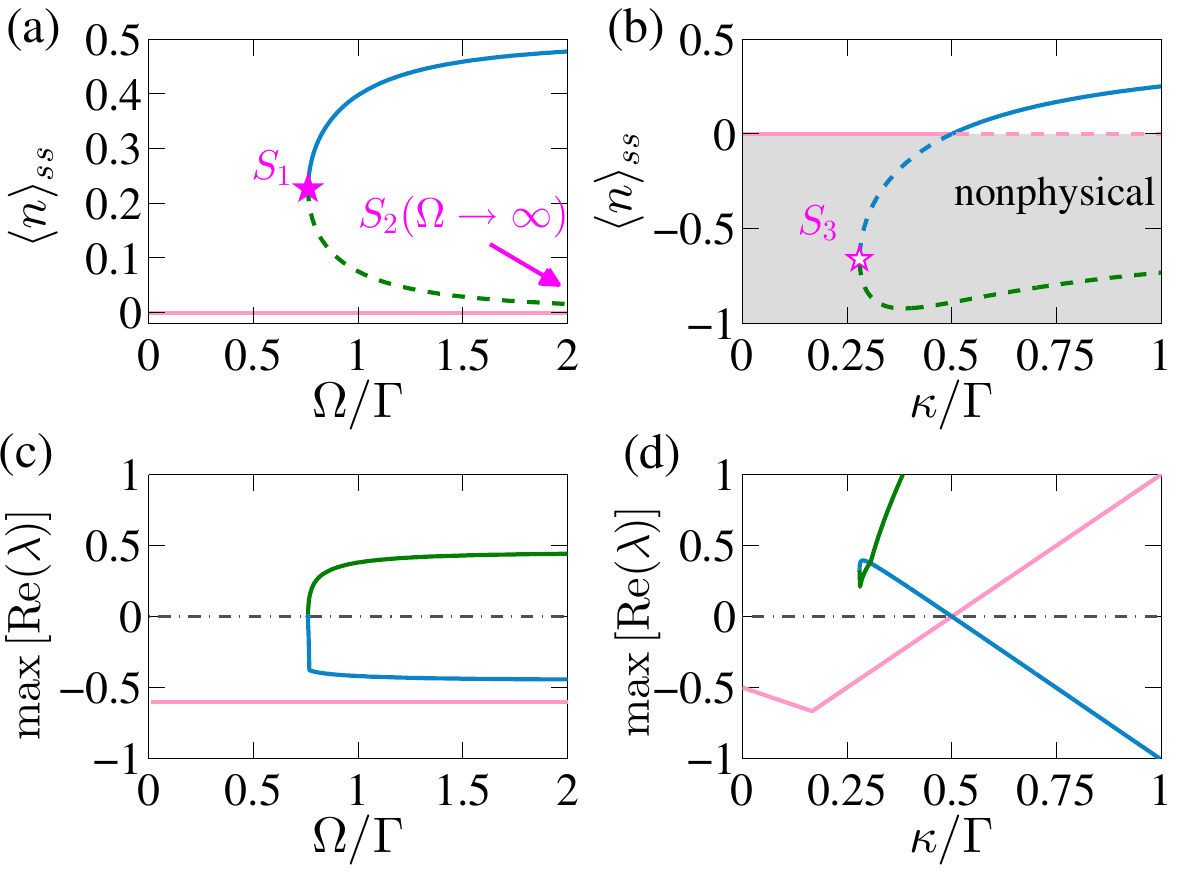}
\caption{The steady-state particle density $\langle \hat{n}\rangle_{\text{ss}}$ for $\kappa/\Gamma=0.1$ (a) and $\Omega/\Gamma=0.1$ (b). The shaded region in gray denotes the nonphysical solutions. The solid lines denote the stable steady-state solution and the dashed lines denote the unstable ones. The two branches of active states, associated with the solution $\langle\hat{\bm{\sigma}}\rangle_{\text{ss},\pm}$, emerges at $S_1$ and $S_3$. The former is a saddle-node point of the bifurcation (with the $S_2$ saddle node locating at infinite $\Omega$), leading to a discontinuous phase transition in panel (a). The intersection of the absorbing and upper active branch at $\kappa/\Gamma=1/2$ indicates a continuous phase transition in panel (b). The maximal real parts of the eigenvalues of the corresponding Jacobian matrix of $\kappa/\Gamma=0.1$ (c) and $\Omega/\Gamma=0.1$ (d). The pink, blue and green curves denote the quantities associated with the steady-state solution $\langle\hat{\bm{\sigma}}\rangle_{\text{ss},0}$ and $\langle\hat{\bm{\sigma}}\rangle_{\text{ss},\pm}$, respectively.
} 
\label{fig_MF_orderparameter}
\end{figure}

Now we are in the position to investigate the continuity of the transitions from the absorbing to the other phases. In order to examine the transition from the absorbing phase to the bistable region, we make a vertical cut along $\kappa/\Gamma=0.1$ in the phase diagram, as indicated by the vertical arrow in Fig. \ref{fig:0}(a). The steady-state particle density as a function of $\Omega/\Gamma$ is shown in Fig. \ref{fig_MF_orderparameter}(a). One finds that all the three branches of $\langle\hat{n}\rangle_{\text{ss}}$ are physical acceptable by means of $\langle \hat{n}\rangle_{\text{ss}}\in[0,1]$. The two branches of active state emerge at the critical point $\Omega_c/\Gamma\approx0.7615$. Note that the upper or lower branch is associated with the $\langle\hat{\bm{\sigma}}\rangle_{\text{ss},\pm}$, respectively. 

The maximal real parts of the eigenvalues of the corresponding Jacobian matrix are shown in \ref{fig_MF_orderparameter}(c). One can see that the absorbing state and the upper branch of the active state are stable, while the lower brach is unstable. This features a saddle-node bifurcation with one saddle-node point at $S_1$ and the other one $S_2$ at infinite $\Omega$. The saddle-node-like bifurcation has been observed in the QCP ~\cite{Shang2026}. The existence of the unstable branch leads to the discontinuous absorbing phase transition.

On the other hand, we make a horizontal cut along $\Omega/\Gamma=0.1$ in the phase diagram Fig. \ref{fig:0}(a) to investigate the transition from the absorbing to active phases. The steady-state particle density and the maximal real parts of the eigenvalues of the corresponding Jacobian matrix are shown in Figs. \ref{fig_MF_orderparameter}(b) and (d). Here in order to get a full picture of all the real steady-state solutions, we have also included the nonphysical (negative) ones. It can be seen that a pair of real solutions $\langle\hat{\bm{\sigma}}\rangle_{\text{ss},\pm}$ emerges at $S_3$, although they are neither physical nor stable. 

However, as $\kappa$ increases, the value of $\langle n\rangle_{\text{ss}}$ in the upper branch keeps increasing and intersects with the absorbing solution $\langle n\rangle_{\text{ss}}=0$ at $\kappa/\Gamma=1/2$. Moreover, while crossing the $\kappa/\Gamma=1/2$ the absorbing solution becomes unstable and the upper branch of the active state becomes physical acceptable and stable, as revealed by the maximal real parts of the Jacobian matrix in Fig. \ref{fig_MF_orderparameter}(d). In the meantime, the lower branch remains unstable over the entire parameter space. Therefore, one can conclude that the transition from the absorbing to the active phases is continuous. We emphasize that such continuous phase transition does not originate from a pitchfork bifurcation of the order parameter. This is essentially different from the continuous steady-state phase transitions with spontaneous symmetry breaking in the dissipative $XYZ$ model ~\cite{Jin2016} and the dissipative transverse-field Ising models~\cite{Jin2018}.

\section{Liouvillian spectrum}
\label{Liouvillian spectrum}
In this section, we characterize the relaxation dynamics of the model
through the spectrum of the Liouvillian superoperator. The Liouvillian
eigenvalues determine the decay rates and oscillation frequencies of the
dynamical modes and, in particular, set the characteristic time scale for
the approach to the stationary state. Consequently, the low-lying
Liouvillian spectrum is also closely related to the computational cost of
simulating the long-time dynamics of open quantum systems
\cite{Minganti2018,Mori2020,Shirai2023,Xie2025}.

To construct a matrix representation of the Liouvillian superoperator
${\cal L}$, we vectorize the density matrix by stacking its columns into
the vector \(\lvert\rho\rangle\rangle\). The arbitrary
operators \(X\) and \(Y\) acting on $\rho$ can be recast as 
\begin{equation}
    \lvert X\rho Y\rangle\rangle=\left(Y^{T}\otimes X\right)
    \lvert\rho\rangle\rangle,
\end{equation}
where the superscript `$T$' denotes the matrix transpose. Under this convention, the Liouvillian superoperator $\cal L$ in Eq. (\ref{eq:(2)}) can be represented as follows,
\begin{eqnarray}
 {\cal L}&=&-i\left(\hat{I}\otimes\hat{H}-\hat{H}^T\otimes\hat{I}\right)\cr\cr
  &&+\sum_{j,\alpha=\pm}{\hat{L}_j^\alpha\otimes(\hat{L}_j^{\alpha})^*-\frac{1}{2}(\hat{L}_j^{\alpha})^\dagger\hat{L}_j^\alpha\otimes\hat{I}-\frac{1}{2}\hat{I}\otimes(\hat{L}_j^{\alpha})^T(\hat{L}_j^{\alpha})^*}\cr\cr
  &&+\sum_{j}{\hat{L}_j\otimes\hat{L}_j^*-\frac{1}{2}\hat{L}_j^\dagger\hat{L}_j\otimes\hat{I}-\frac{1}{2}\hat{I}\otimes\hat{L}_j^T\hat{L}_j^*},
\label{eq:liouvillian_matrix}
\end{eqnarray}
where $\hat{I}$ denotes the identity operator and the superscripts `$*$', and `$\dagger$' denote the matrix complex and Hermitian conjugation, respectively. The Liouvillian superoperator $\cal{L}$ satisfies the following eigenvalue equation,
\begin{equation}
    {\cal L}|r_{i}\rangle\rangle=\mu_{i}|r_{i}\rangle\rangle,
\end{equation}
where $\mu_{i}$ and $|r_{i}\rangle\rangle$ are the Liouvillian
eigenvalues and the corresponding right eigenvectors.

In general, the eigenvalues $\mu_i$ are complex because ${\cal L}$ is non-Hermitian. The real parts of all the eigenvalues satisfy $\operatorname{Re}(\mu_{i})\le0$, and hence describe decaying dynamical
modes. The steady state
is given by the eigenstate associated with the zero eigenvalue,
\begin{equation}
    {\cal L}|\rho_{\mathrm{ss}}\rangle\rangle=0.
\end{equation}
The corresponding right eigenvector can be reshaped into the steady-state
density matrix and normalized according to
$\operatorname{Tr}(\rho_{\mathrm{ss}})=1$. The slowest relaxation mode is determined by the nonzero eigenvalue with the largest (negative) real part. Accordingly, the Liouvillian gap is defined as $\mu_0 = \left|\max{[\operatorname{Re}(\mu_i)]}\right|$. A smaller $\mu_0$ implies a longer relaxation time. The closing of the Liouvillian gap in the thermodynamic limit is commonly associated with the emergence of bistability or a dissipative phase transition~\cite{Minganti2018,Fazio2025}.
\begin{figure}[t!]
\centering
\includegraphics[width=1\linewidth]{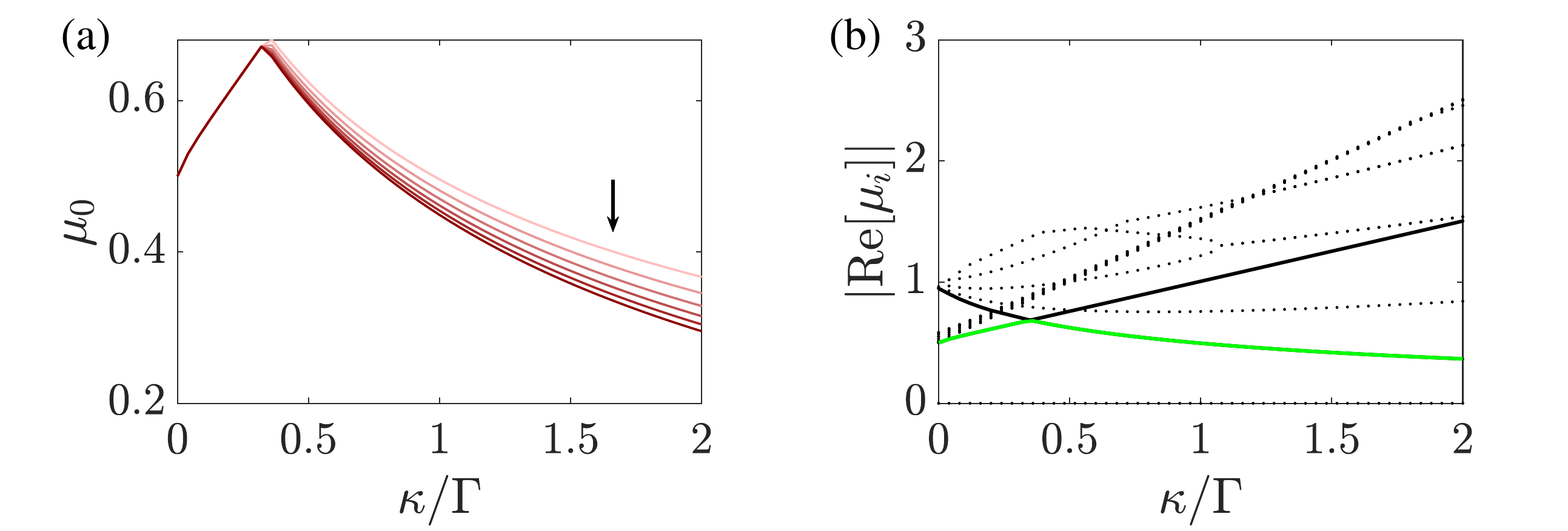}
\caption{(a) The Liouvillian gap of finite-size systems as a function of $\kappa/\Gamma$ for $\Omega/\Gamma = 0.1$, with the arrow indicating the change in the system size $L$ from $7$ to $12$. (b) Displays the real part of the low-lying Liouvillian eigenvalues as a function of $\kappa/\Gamma$ with $L = 7$ for $\Omega/\Gamma = 0.1$. The two solid lines denote the (nearly) linear increasing branch and the monotonically decreasing branch, the minimum of which yields the Liouvillian gap, as highlighted in green.
} 
\label{fig:3}
\end{figure}

We first examine Liouvillian gap for the finite-size system. In Fig. \ref{fig:3}(a), it is shown the $\mu_0$ as a function of $\kappa/\Gamma$ for various system sizes ranging from $L=7$ to $12$, for $\Omega/\Gamma=0.1$. One can see that the Liouvillian gap decreases gradually with the system size increasing over the considered parameter regime. This trend indicates a progressive slowing down of the relaxation dynamics and is consistent with a possible gap closing in the thermodynamic limit.

Moreover, the Liouvillian gap exhibits a nonmonotonic behavior, there is a turning point around $\kappa/\Gamma\approx0.35$. To identify the spectral modes responsible for this behavior, we diagonalize the ${\cal L}$ for $L=7$ and show the first twenty (degenerate) $|\text{Re}(\mu_{i})|$ as functions of $\kappa/\Gamma$ in Fig. \ref{fig:3}(b). One can observe that, as $\kappa$ increases, one branch grows almost linearly and the other one decreases monotonically. They intersect at a turning point as indicated by the two solid lines in Fig. \ref{fig:3}(b). The Liouvillian gap is basically determined by the minimum of these two branches.

The behavior of the Lioullian spectrum of the HCP is different from that of the QCP as reported in Ref. \cite{Shang2026}. In the QCP, the eigenvalue with the largest (negative) nonzero real part is separated from the rest nonzero eigenvalues which are bounded by $-1/2$. The isolation of the non-negative eigenvalues with the first two largest real parts from the rest ones implies the metastability of the time evolution. That is the system may stay in the metastable state for finitely long time before reaching the eventual steady state. However, here no evidence of the metastability in the HCP is observed.

\section{Cluster Mean-field approximation}
\label{Cluster Mean-field approximation}

\begin{figure*}[t]
    \centering
    \includegraphics[width=0.9\textwidth]{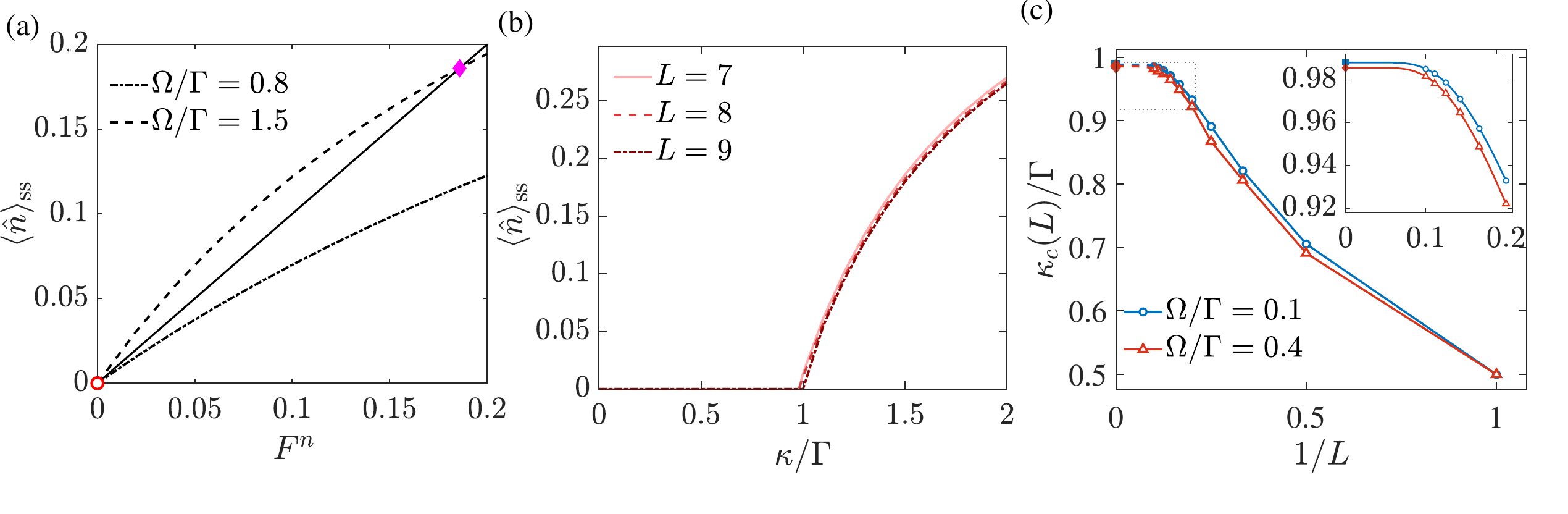}
    \caption{
(a) The $\langle\hat{n}\rangle_{\mathrm{ss}}$ as a
function of the effective field $F^n$ for $L=7$ and
$\Omega/\Gamma=0.1$. The black solid line represents the self-consistency
condition $\langle\hat{n}\rangle_{\mathrm{ss}}=F^n$.
The dashed and dash-dotted curves show the $\langle\hat{n}\rangle_{\mathrm{ss}}$ for
$\kappa/\Gamma=1.5$ and $0.8$, respectively. Their intersection marks the self-consistent steady-state
solution. The nonzero and zero solutions are indicated by the filled
magenta diamond and the empty red circle, respectively.
(b) The $\langle\hat{n}\rangle_{\mathrm{ss}}$ as a
function of $\kappa/\Gamma$ for $\Omega/\Gamma=0.1$, with $L=7$ , $8$, and $9$.
(c) The CMF critical point
$\kappa_c(L)/\Gamma$ against the  $1/L$ for $\Omega/\Gamma=0.1$ and $0.4$. The inset provides an enlarged view of the region
used for the exponential fits and thermodynamic limit extrapolation ($1/L\rightarrow 0$). The filled symbols at $1/L=0$ indicate the extrapolated critical
points.The extrapolated critical points are $\kappa/\Gamma=0.9881$ and $0.9857$ for $\Omega/\Gamma=0.1$ and $0.4$, respectively.
}
    \label{fig:4}
\end{figure*}

In the MF approximation, all the correlations are neglected in the analysis. Now we are going to incorporate the short-range correlations by utilizing the CMF method. The idea of CMF is to extend the Gutzwiller factorization of the total density matrix to a tensor product of the density matrices of identical clusters as follows,
\begin{equation}
\rho=\underset{C}{\bigotimes}\rho_C,
\label{CMF_dm}
\end{equation}
where $\rho_C$ represents the density matrix of the cluster, which is composed of a number of linked sites. The single-site approximation can thus be regarded as a limit case of the CMF. In the CMF approximation, the interactions inside the cluster are treated in a full quantum way. Therefore, the correlations among the sites could be gradually included as the size increases. In principle, as the size of the cluster increases larger and larger, the CMF result should converge to the true one. 

Since the clusters are assumed to be identical, we use $\rho_C$ to denote the density matrix of arbitrary cluster. 
Substituting Eq. (\ref{CMF_dm}) into Eq.~\eqref{eq:(2)} and taking partial trace over the rest clusters, one can obtain the master equation for the density matrix of a certain cluster $C$,
\begin{equation}
\frac{\text{d}}{\text{d}t}\rho_{C}=-i[\hat{H}_{\text{CMF}},\rho_C]+{\cal D}_{\text{corr}}^{\text{CMF}}[\rho_C]+ {\cal D}_{\text{loc}}^{\text{CMF}}[\rho_C],
\label{eq:CMF_ME}
\end{equation}
where $\hat{H}_{\text{CMF}}=\hat{H}_C+\hat{H}_{C\text{-}C^{\prime}}$ is the CMF Hamiltonian with
\begin{equation} \hat{H}_{C}=\Omega\sum_j{(}\hat{\sigma}_{j}^{x}\hat{n}_{j+1}+\hat{n}_j\hat{\sigma}_{j+1}^{x})  
\end{equation}
representing the internal Hamiltonian of the cluster, and 
\begin{equation}
\hat{H}_{C\text{-}C^{\prime}}=\Omega\sum_{\langle j,j^{\prime}\rangle}{\left(\langle\hat{\sigma}_{j^{\prime}}^{x}\rangle \hat{n}_j+\langle \hat{n}_{j^{\prime}}\rangle\hat{\sigma}_{j}^{x}\right)}
\label{eq:HCMF_interC}
\end{equation}
describing the interaction between the cluster $C$ and its neighboring cluster $C'$, and $j \in \partial C$, $j' \in \partial C'$ denote the boundary sites of $C$ and $C'$, respectively \cite{Jin2016}.

Similarly, the correlated dissipator in the CMF approximation is given by 
\begin{eqnarray}
{\cal D}_{\text{corr}}^{\text{CMF}}[\rho_C] &=& \kappa\sum_{\langle j,j'\rangle}[\langle \hat{n}_{j'}\rangle(\hat{\sigma}^+_j\rho_C\hat{\sigma}^-_j+\hat{\sigma}_{j}^-\rho_C\hat{\sigma}_{j}^+-\rho_C)\cr\cr&& + \hat{n}_j\rho_C\hat{n}_j  
 - \frac{1}{2}\{\hat{n}_j,\rho_C\}],
\label{eq:corrdiss}
\end{eqnarray}
and the local dissipator is given by
\begin{equation}
{\cal D}_{\text{loc}}^{\text{CMF}}[\rho_C] = \Gamma\sum_{j}\left(\hat{\sigma}^{-}_j\rho_C\hat{\sigma}^+_j -\frac{1}{2} \{\hat{n}_j,\rho_C\}\right).
\end{equation}
Due to the translational invariance of the system, the steady-state solution of the CMF master equation (\ref{eq:CMF_ME}) can be determined by the self-consistent condition $\langle\hat{O}_{j'}(t\rightarrow\infty)\rangle = \text{tr}[\hat{O}_{j}\rho_C(t\rightarrow\infty)]$ for $\hat{O}=\hat{n}$ and $\hat{\sigma}^x$. 

From Eqs. (\ref{eq:HCMF_interC}) and (\ref{eq:corrdiss}), one can see that the interactions between the boundary sites belonging to the neighboring clusters are approximated by the effective fields imposing on the boundary. So one can reexpress Eqs. (\ref{eq:HCMF_interC}) and (\ref{eq:corrdiss}) by replacing the $\langle \hat{n}_{j'}\rangle$ and $\langle\hat{\sigma}^x_{j'}\rangle$ with the effective fields $F^n$ and $F^x$, respectively. In this manner, the steady-state density matrix is parameterized as $\rho_{C,\text{ss}}(F^n,F^x)$. Note that the solution with nonzero $\langle\hat{\sigma}^x\rangle_{\text{ss}}$ is not physically acceptable \cite{Michael2017}, thus only $F^n$ is involved in the parameterization. The self-consistent condition then requires that $\langle\hat{n}\rangle_{\text{ss}}=\text{tr}[\hat{n}_j\rho_{C,\text{ss}}(F^n)]$ for $j=1$ or $L$ (the boundary site). In this way, all possible steady states can be captured even if it is unstable. Therefore, a comprehensive picture of the steady-state solutions can be obtained to uncover the (dis)continuity of the phase transitions. 

In Fig. \ref{fig:4}(a) we compute the steady-state particle number of the boundary site, $\langle\hat{n}_1\rangle_{\text{ss}}$, as a function of $F^n$ for $L=7$. One can see that for $\kappa/\Gamma = 0.8$ the curve of $\langle\hat{n}_1\rangle_{\text{ss}}$ intersects with the line $F^n=\langle\hat{n}_1\rangle_{\text{ss}}$ only at the origin, indicating that the absorbing state is the unique stable solution. By contrast, for $\kappa/\Gamma = 1.5$ an additional intersection appears at $F^n = 0.1859$, corresponding to the existence of an active state. Moreover the active state is stable because the derivative of $\langle\hat{n}\rangle_{\text{ss}}$ at the intersection is less than 1. Figure \ref{fig:4}(b) presents the distribution of steady-state solutions for various $L$. As the short-range correlations are gradually included, the $\langle\hat{n}\rangle_{\text{ss}}$ converges and the feature of continuous phase transition is preserved. In contrast to the QCP, in which the trends of the transition point exhibits an even-odd size effect ~\cite{Shang2026}, here the transition points for different $L$ approach the true critical point from the left side. This can be seen from the dependence of the critical points on the size $L$, as shown in Fig. \ref{fig:4}(c). For both $\Omega/\Gamma=0.1$ and $0.4$, we use the data from $L=5$ to $10$ to extrapolate the critical point in the thermodynamic limit. The data were fitted by the exponential finite-size scaling form  $\kappa_c(L)/\Gamma=\kappa_c/\Gamma - A\exp(-bL)$, where $A$ and $b$ are the fitting parameters and $\kappa_c$ denotes the critical point in the thermodynamic limit. The value of $\kappa_c$ was obtained from the extrapolation $1/L\rightarrow0$, which results in $\kappa_{c}/\Gamma= 0.9881$ and $0.9857$, respectively.

\section{Coherent anomaly method}
\label{Coherent anomaly method}
The critical exponent  is of great significance in classifying the universality of the continuous phase transition. Although the CMF approximation refines the critical point over the MF approximation, the critical exponent still falls into the classical ones because of the MF nature of the method. However, the coherent anomaly method (CAM), first proposed by Suzuki, provides a way to reveal the true critical exponent of the nonclassical phase transition by systematically analyzing a series of CMF results ~\cite{Suzuki1986,Suzuki1987,Park2005}. One of the authors of the present paper has generalized the CAM to the dissipative phase transition \cite{Jin2021}.

Here we review the idea of CAM briefly. At the vicinity of the phase transition, the order parameter in a CMF approximation with size $L$ scales as follows,
\begin{equation}
\langle\hat{n}(\kappa,L)\rangle_{\text{ss}}\approx n_0(L) \left[\frac{\kappa_c(L)-\kappa}{\kappa_c(L)}\right]^{\beta^{\text{mf}}},
\end{equation}
where $\kappa_c(L)$ is the critical point of the CMF approximation with size $L$, $\beta^{mf}$ is the classical critical exponent, and $n_0(L)$ is the amplitude which is also referred to as the coherent anomaly. As the size $L$ varies, the scaling of $\langle\hat{n}(\kappa,L)\rangle_{\text{ss}}$ remains characterized by the classical exponent $\beta^{\text{mf}}$, which is usually known. For example, for the classical contact process the classical critical exponent of the order parameter is $\beta^{\text{mf}}=1$. However, for various $L$, the CMF critical point $\kappa_c(L)$ shifts and the value of the order parameter differs due to the coherent anomaly $n_0(L)$, which scales as follows,
\begin{equation}
n_0(L)\sim C_0\left[\frac{\kappa^*-\kappa_c(L)}{\kappa^*}\right]^{-(\beta^*-\beta^{\text{mf}})},
\label{eq:CAM}
\end{equation}
where $\kappa^*$ and $\beta^*$ are the true critical point and critical exponent, and $C_0$ is the coefficient to be determined. A detailed derivation of Eq. (\ref{eq:CAM}) can be found in Refs. \cite{Suzuki1987,Jin2021}.
 
\begin{figure}[htbp]
\centering
\includegraphics[width=1\linewidth]{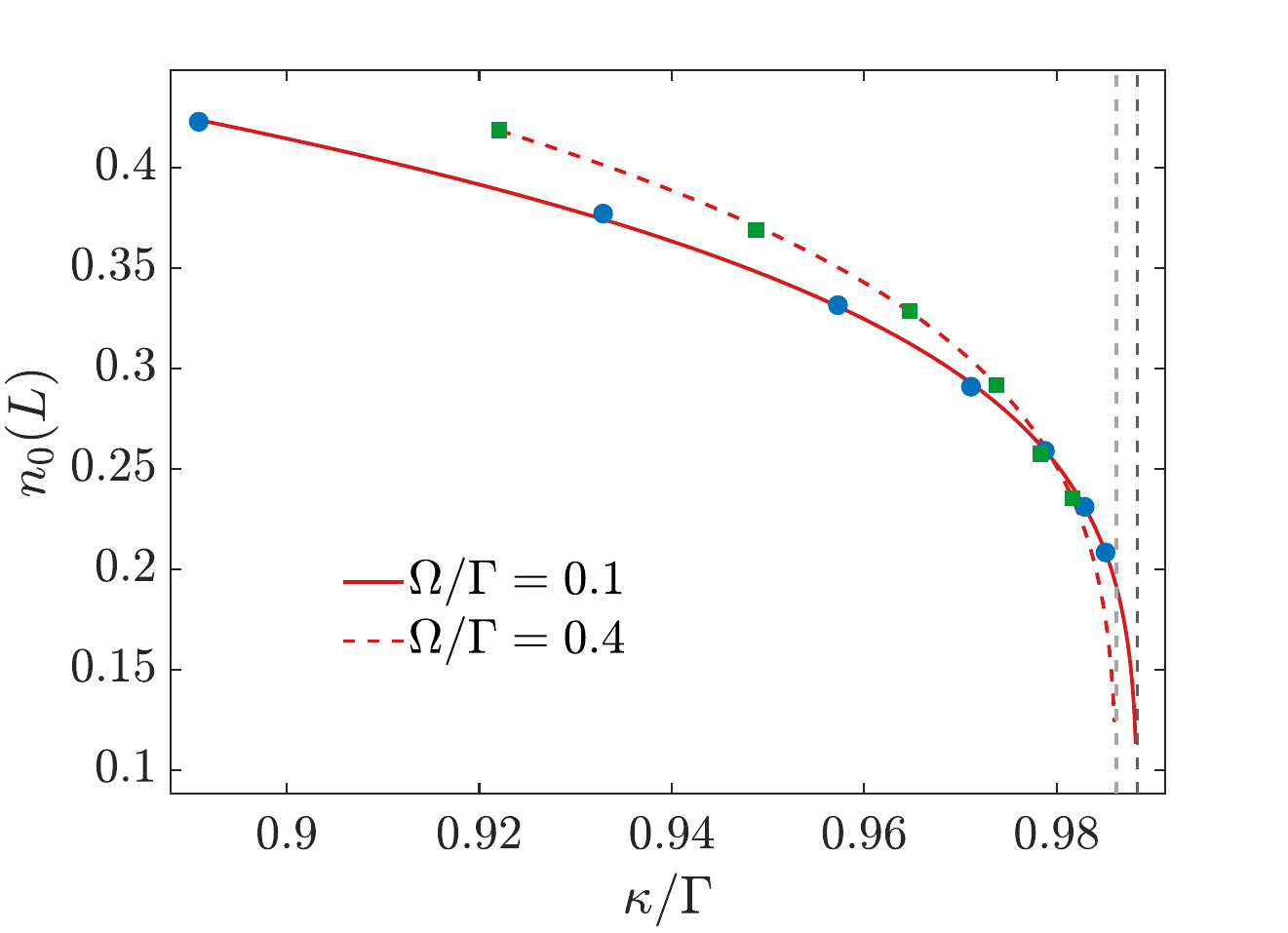}
\caption{The CAM analysis of the cluster mean-field results. The blue circles and green squares denote the coherent anomalies $n_0(L)$ for $\Omega/\Gamma=0.1$ and $0.4$, respectively. The solid and dashed lines in red are the corresponding fittings with Eq. (\ref{eq:CAM}). For $\Omega/\Gamma=0.1$, the coherent anomalies with
$L=5,6,..., 10$ are used for the fitting, while for $\Omega/\Gamma=0.4$, the coherent anomalies with
$L=6,7,...,10$ are used. 
The vertical dashed lines in dark- and light-gray mark the estimated critical points $\kappa^*/\Gamma=0.9884$ and $0.9862$ for $\Omega/\Gamma=0.1$ and $\Omega/\Gamma=0.4$, respectively. The corresponding critical exponents are estimated as $\beta^*=0.7962$ and $0.7865$, for $\Omega/\Gamma=0.1$ and $0.4$, respectively.} 
\label{fig:5}
\end{figure}

With the help of Eq. (\ref{eq:CAM}), we can extract the true critical point and exponent through a series of CMF results $\{n_0(L),\kappa_c(L)\}$. Figure~\ref{fig:5} presents the fitting of $n_0(L)$ with $L=5,6,...,10$. The critical points obtained from the CAM are $\kappa^{*}/\Gamma=0.9884$ and $0.9862$ for $\Omega/\Gamma=0.1$ and $0.4$, respectively. The CAM critical points agree with those from the direct extrapolation of $\kappa_c(L)$ with $1/L$, as shown in Fig.~\ref{fig:4}(c). The true critical exponents are estimated as $\beta^*=0.7962$ and $0.7865$ for $\Omega/\Gamma=0.1$ and $0.4$, respectively. The critical exponents $\beta^*$ of the continuous phase transition in the HCP for various $\Omega$ seem to be very close to each other. We would emphasize that CMF results with larger sizes are required to obtain a more precise $\beta^*$.

\section{Summary}
\label{summary}
In summary, we have investigated the nonequilibrium steady-state phases and critical behavior of the one-dimensional HCP, in which coherent and incoherent branching and coagulation coexist with local decay. By means of the single-site MF approximation, we obtain a phase diagram consisting of absorbing phase, active phase, and bistable regions. As the coherent coupling $\Omega$ varies, the transition from the absorbing phase to the bistable region is accompanied by a saddle-node bifurcation, resulting in a discontinuous phase transition. In contrast, as the correlated decay rate $\kappa$ varies, the continuous transition from the absorbing to active phases occurs. It should be noted that such continuous transition does not originate from the pitchfork bifurcation associated with spontaneous symmetry breaking.

We have further characterized the relaxation dynamics through the low-lying Liouvillian spectrum. The Liouvillian gap decreases progressively with increasing system size and exhibits a nonmonotonic dependence on the correlated decay rate $\kappa$ due to the crossing of two low-lying spectral branches. Although its closing in the thermodynamic limit cannot be resolved with the accessible system sizes, the observed finite-size behavior indicates critical slowing down. Unlike the QCP, no isolated slow mode or clear signature of metastability is found in the HCP.

To incorporate short-range correlations, we have employed the CMF together with a self-consistent effective-field condition that captures both stable and unstable steady-state solutions. The continuous character of the absorbing-to-active transition is preserved as the cluster size increases. The extrapolated cluster critical points agree well with those obtained from the CAM. The critical exponents $\beta^*$ of the order parameter are extracted with CAM for different $\Omega$. The estimated $\beta^*$ with the accessible system sizes suggests that the continuous transitions may share the same critical behavior. Calculations with larger clusters are nevertheless required to obtain more accurate critical exponents and clarify the universality of the HCP.

\begin{acknowledgments}
This work is supported by the Natural Science Foundation of Liaoning Province No. 2025-MS-009. X. L. is supported by the National Natural Science Foundation of China (Grant No. 12247101), the Fundamental Research Funds for the Central Universities (Grant No. lzujbky-2025-jdzx07), the Natural Science Foundation of Gansu Province (No.25JRRA799), and the ‘111 Center’ under Grant No. B20063.
\end{acknowledgments}

\nocite{apsrev42Control}
\bibliographystyle{apsrev4-2}
\bibliography{bibliography}

\end{document}